\documentclass[12pt]{article}
\usepackage{graphicx}

\begin{document}

\author{
T. M. Aliev\thanks{taliev@metu.edu.tr},
~ K. Azizi\thanks{e146342@metu.edu.tr},
~ A. Ozpineci\thanks{ozpineci@metu.edu.tr},
~ M. Savci\thanks{savci@metu.edu.tr} \\ \small
Physics Dept., Middle East Technical University, 06531, Ankara, Turkey}

\title{ { Nucleon Electromagnetic Form Factors in QCD } }
\begin{titlepage}
\maketitle
\begin{abstract}
 The nucleon electromagnetic form factors  are calculated  in light cone QCD sum rules framework using the most general form of the
nucleon interpolating current.  Using two forms of
the distribution amplitudes (DA's),   predictions for the form factors are presented
and compared  with existing experimental data. It is shown  that
our results describe remarkably well the existing experimental data.
\end{abstract}
\thispagestyle{empty}
\end{titlepage}
\section{Introduction}
The nucleon electromagnetic (EM) form factors are the fundamental objects for
 understanding  their internal structure. The internal structures
of the nucleon are usually described in terms of the electromagnetic
Dirac and Pauli form factors $F_{1}(q^{2})$ and $F_{2}(q^{2})$ or
equivalently the electric and magnetic dipole Sachs form factors
$G_{E}(q^{2})$ and
 $G_{M}(q^{2})$, respectively (for a recent status of experiments and phenomenology of the form factors see \cite{Arringtin}).

Until a few years ago, the nucleon electromagnetic form factors are
studied in unpolarized elastic electron-nucleon scattering through a
virtual photon exchange. It is shown in the pioneering work \cite{Akhiezer} that the polarization effects, i.e., scattering of
polarized electrons from polarized target, can play essential role
for a more accurate determination of the nucleon electromagnetic form
factors. The main result of \cite{Akhiezer} is that, unlike the
unpolarized elastic cross section, which is proportional to the sum
of squares of the form factors, the polarized cross section contains
also interference terms of the form factors $G_{E}(q^{2})$ and
 $G_{M}(q^{2})$. Studying various polarization observables allows more accurate determination of these form factors.

Recent developments in experimental instruments allow to produce polarized electron beams and polarized protons,
which gives the opportunity for a  more precise separation of the  $G_{E}(q^{2})$ and
 $G_{M}(q^{2})$ form factors. The electron-proton scattering experiments, 
 which are performed at Jefferson Laboratory using the polarized electrons and polarized proton, show strong deviation from the theoretical predictions \cite{Jones,Gayou,Gayou2,Punjabi},
  i.e., the ratio $F_{2}(q^{2})/F_{1}(q^{2})$ does not behave as is expected from previous experiments and as is predicted
  by the perturbative QCD (for review see \cite{Perdrisat} and references therein). For understanding this unexpected result,
  some model-independent non perturbative method is needed. Among all existing nonperturbative approaches QCD sum rule is
  more attractive and powerful, because it is based on the fundamental QCD Lagrangian.

The goal of our work is the calculation of the electromagnetic form
factors of nucleon using the light cone QCD sum rule (LCQSR)
 and most general form of the interpolating current for nucleon. In this approach the form factors of the nucleons are expressed in
 terms of distribution amplitude of the nucleon. Note that, this problem is investigated for the Ioffe current in the framework of
 the LCQSR in \cite{Lenz} and the traditional sum rules in \cite{Castillo}. In \cite{Braun2},  an improved version of the Chernyak–Zhitnitsky current is used.
 The paper is organized in following way. in section 2 , we present the result for the nucleon electromagnetic form factors in the LCQSR method. Section 3 is devoted to the numerical analysis, discussion and conclusion.

\section{ Electromagnetic form factors of  nucleon in LCQSR }
In this section EM form factors of nucleon are
calculated within the light cone QCD sum rules method. The
electromagnetic form factors of nucleon are defined by the matrix
element of the electromagnetic current $J^{el}_{\lambda}$ between
the initial and final nucleon states $\langle N(p')\mid
J^{el}_{\lambda}\mid N(p)\rangle$. The most general form of this matrix
element satisfying the Lorentz invariance and electromagnetic
current conservation is
\begin{eqnarray}\label{matrixel1}
\langle N(p')\mid
J^{el}_{\lambda}(0)\mid N(p)\rangle=\bar N(p')\left[\gamma_{\lambda}F_{1}(Q^{2})-\frac{i}{2m_{N}}\sigma_{\lambda\nu}q^{\nu}F_{2}(Q^{2})
\vphantom{\int_0^{x_2}}\right]N(p),\nonumber\\
\end{eqnarray}
where $Q^{2}=-q^{2}$, is the negative of the square of the virtual
photon momentum, $q=p-p'$ and $F_{1}$ and $F_{2}$ are the Dirac and
Pauli form factors, respectively. 

Another set of nucleon form factors is the so called Sachs form factors,
which are defined in terms of the $F_{1}(Q^{2})$ and $F_{2}(Q^{2})$
as follows:
\begin{eqnarray}
&&G_{M}(Q^{2})=F_{1}(Q^{2})+F_{2}(Q^{2}),\nonumber\\
&&G_{E}(Q^{2})=F_{1}(Q^{2})-\frac{Q^{2}}{4 m_{N}^{2}}F_{2}(Q^{2}),
\end{eqnarray}
At the static limit, values at  $Q^{2}=0$ are  $G_{E}^{P}(0)=1$,
$G_{E}^{n}(0)=0$, $G_{M}^{P}(0)=\mu_{P}= 2.792847337(29)$
 and $G_{M}^{n}(0)=\mu_{n}= -1.91304272(45)$, where $\mu_{P}$ and $\mu_{n}$ are the anomalous magnetic moments of the proton and neutron in units of the Bohr magneton.\\

 After these preliminary remarks, we proceed to  calculate the
 electromagnetic form factors of nucleon in LCQSR.
 The basic object of the LCQSR is a suitably chosen correlation function. In this study, it is chosen as:
\begin{equation}\label{T}
\Pi_{\lambda}(p,q)=i\int d^{4}xe^{iqx}\langle0\mid
T\{J^{N}(0)J^{el}_{\lambda}(x) \}\mid N(p)\rangle,
\end{equation}
which describes the transition of the nucleon $N(p)$ to the nucleon $N(p-q)$ via the EM current. The interpolating current for the nucleon is  chosen as
 \begin{equation}\label{cur.N}
J^{N}(x)=2\varepsilon^{abc}\sum_{\ell=1}^{2}(u^{T a} (x)CA_{1}^{\ell} d^{b} (x) )A_{2}^{\ell} u^{c} (x),
\end{equation}
where $A_{1}^{1}=I$, $A_{1}^{2}=A_{2}^{1}=\gamma_{5}$, $A_{2}^{2}=\beta$, and  C is the charge conjugation operator, and a, b, c are the color indices.
  The electromagnetic  current is:
\begin{equation}\label{cur.elect}
J^{el}_{\lambda}(x)=e_{u}\bar u\gamma_{\lambda}u+e_{d}\bar d\gamma_{\lambda}d,
\end{equation}
and the choice $\beta=-1$ corresponds to the Ioffe current.
The main idea of the LCQSR method is to calculate the correlation
function in terms of the form factors   at hadron level,  as
well as in terms of the quark and gluon degrees of freedom. Equating  two representations of the
correlation function and performing a Borel transformation in order
to suppress the contributions of the higher states and continuum, we
get sum rules for the EM form factors of the nucleon.

Let us first calculate the physical part of the correlator (\ref{T}). The contribution of the nucleon to the correlation function (\ref{T}) is given by
\begin{equation}\label{phys1}
\Pi_{\lambda}(p,q)=\sum_{s}\frac{\langle0\mid
J^{N}(0)\mid N(p',s)\rangle\langle N(p',s)\mid
J^{el}_{\lambda}(0)\mid N(p)\rangle}{m_{N}^{2}-p'^{2}}.
\end{equation}
The matrix element $\langle0\mid
J^{N}(0)\mid N(p',s)\rangle$  in (\ref{phys1}) is determined in the following way:
\begin{equation}\label{matrixel2}
\langle0\mid
J^{N}(0)\mid N(p',s)\rangle=\lambda_{N} N(p',s),
\end{equation}
where $\lambda_{N}$ is the coupling constant of the nucleon to the
current $J^{N}(0)$. The matrix element $\langle N(p',s)\mid
J^{el}_{\lambda}(0)\mid N(p)\rangle$ is parameterized in terms of
the form factors $F_{1}$ and $F_{2}$ via Eq. (\ref{matrixel1}).
Summing over spins of the nucleons 
\begin{equation}\label{spinor}
\sum_{s}N(p',s)\overline{N}(p',s)=\not\!p'+m_{N},
\end{equation}
and using Eqs.
(\ref{matrixel1}), (\ref{phys1}) and (\ref{matrixel2}), we obtain
the following expression for the contribution of nucleon to the
correlation function
\begin{eqnarray}\label{phys2}
\Pi_{\lambda}(p,q)=\frac{\lambda_{N}}{m_{N}^{2}-p'^{2}}(\not\!p'+m_{N})\left[\gamma_{\lambda}F_{1}(Q^{2})-\frac{i}{2m_{N}}\sigma_{\lambda\nu}q^{\nu}F_{2}(Q^{2})\vphantom{\int_0^{x_2}}\right]N(p)
+ \cdots \nonumber\\
\end{eqnarray}
where $\cdots$ stand for the contributions to the correlation functions from the higher states and continuum.
It follows from expression (\ref{phys2}) that,  the correlation function contains numerous structures
and in principle all of them can be used in determination of the electromagnetic form factors of nucleons. In further analysis, we choose the independent structures containing $p_{\lambda}$,
 and $p_{\lambda}\!\!\not\!q$ for obtaining $F_{1}$ and $F_{2}$, respectively.

The theoretical part of the  correlator can be calculated in LCQSR in
deep
 Euclidean region
$p'^{2}=(p-q)^{2} << 0$ in terms of the nucleon DA's. These nucleon DA's for all three quarks have been studied in great detail in  \cite{Lenz, Braun1, Braun2}.
 Using the explicit expression for the currents and carrying out all contractions, 
 the correlation function takes the form 
\begin{eqnarray}\label{mut.m}
\left(\Pi_\lambda\right)_\rho &=& \frac{i}{2} \int d^4x e^{iqx} \sum_{\ell=1}^2 \Bigg\{\nonumber\\&&
e_{u}( C A_1^\ell)_{\alpha\gamma} \Big[ A_2^\ell S_u(-x) \gamma_\lambda
\Big]_{\rho\phi}
4 \epsilon^{abc} \langle 0 | u_\alpha^a(0) u_\phi^b(x)
d_\gamma^c(0) | N(p) \rangle \nonumber\\
&+&e_{u}( A_2^\ell )_{\rho\alpha} \Big[ ( C A_1^\ell)^T S_u(-x)
\gamma_\lambda\Big]_{\gamma\phi} 4 \epsilon^{abc} \langle 0 |
u_\alpha^a(0)
u_\phi^b(x) d_\gamma^c(0) | N(p) \rangle\nonumber \\
&+& e_{d}( A_2^\ell )_{\rho\phi} \Big[ C A_1^\ell S_d(-x)
\gamma_\lambda\Big]_{\alpha\gamma} 4 \epsilon^{abc} \langle 0 | u_\alpha^a(0)
u_\phi^b(0)  d_\gamma^c(x) | N (p)\rangle \Bigg\},\nonumber\\
\end{eqnarray}
in $x$ representation, where $\lambda$ is a Lorentz index, and
$\alpha,~\gamma,~\rho$ and $\phi$ are spinor indices.
 $ S(x)$ is  the light cone expanded light quark full propagator
 \cite{Balitsky} having the form:
\begin{eqnarray}\label{prop}
S(x) &=& \frac{i\not\!x}{2\pi^2x^4}
-<qq>(1+\frac{m_{0}^{2}x^{2}}{16})-ig_{s} \int\limits_0^1dv
[\frac{\not\!x}{16\pi^2x^2}G_{\mu\nu}\sigma^{\mu\nu}\nonumber\\
&-&vx^{\mu}G_{\mu\nu}\gamma^{\nu}\frac{i}{4\pi^2x^2}],
\end{eqnarray}
where $m_0^2 = (0.8 \pm 0.2)~GeV^2$ and $G_{\mu \nu}$ is the gluon field strength tensor.
The terms proportional to the gluon strength tensor can give
contribution to four- and five-particle distribution functions but
they are expected to be very small  \cite{Lenz, Braun1, Braun2} and for this reason we
will neglect these amplitudes in further analysis. The terms
proportional to $<qq>$ can also be omitted because Borel
transformation eliminates these terms and hence only the first term in
Eq. (\ref{prop}) is relevant for our discussion. It follows from Eq.
(\ref{mut.m})  that for the calculation of
$\Pi_{\lambda}(p,q)$ we need to know the matrix element\\
\begin{equation}\label{uud1}
 \langle0\mid
4\epsilon^{abc}u_{\alpha}^{a}(a_{1}x)u_{\phi}^{b}(a_{2}x)d_{\gamma}^{c}(a_{3}x)\mid N(p)\rangle.
\end{equation}
It is shown in \cite{Braun1} that the general Lorentz
decomposition of this matrix element is symmetric with
respect to interchange of the momentum fractions of the u-quarks:
\begin{equation}\label{uud2}
 \langle0\mid
4\epsilon^{abc}u_{\alpha}^{a}(a_{1}x)u_{\phi}^{b}(a_{2}x)d_{\gamma}^{c}(a_{3}x)\mid
N(p)\rangle=\sum K \Gamma_{1}^{\alpha\phi}\left( \Gamma_{2} N(p) \right)^{\gamma},
\end{equation}
where $N(p)$ on the right is the nucleon spinor, $\Gamma_{1,2}$ are certain Dirac structures over which the sum is carried out, 
$a_{i}$ are positive numbers which satisfy $a_{1}+a_{2}+a_{3}=1$, and $K$
 are the distribution amplitudes, depending on eight nonperturbative parameters. Explicit expressions of all DA'S and the values of
 eight nonperturbative parameters can be found in \cite{Lenz, Braun1, Braun2, Aliev1}.

Omitting the details of calculations of the theoretical part,
 choosing the coefficients of the structures $p_{\lambda}$,
 and $p_{\lambda}\!\!\not\!q$, equating both representation of the correlation function
 and applying the Borel transformation with respect to the variable
 $p'^{2}=(p-q)^{2}$, which suppress the contributions of the higher
 states and continuum, we obtain following sum rules for the form factors $F_{1}$ and
 $F_{2}$:
\begin{eqnarray}\label{F_{1}}
&&F_{1}(Q^{2})=
\frac{-1}{2\lambda_{N}}  e^{m_{N}^{2}/M_{B}^{2}}\left\{e_{u}m_{N}\int_{t_{0}}^{1}dx_{2}\int_{0}^{1-x_{2}}dx_{1}
e^{-s(x_{2},Q^{2})/M_{B}^{2}}\right[2{\cal H}_{5,-7}(x_i)(1-\beta)
\nonumber \\ &&+4({\cal H}_{17}(x_i) -2{\cal H}_{19}(x_i))(1+\beta)\vphantom{\int_0^{x_2}}\left]
+e_{u}m_{N}\int_{t_0}^1dx_2\int_0^{1-x_2}dx_1\int_{t_0}^{x_2}\frac{dt_1}{t_1}e^{-s(t_1,Q^2)/M_{B}^{2}
}\right(
\nonumber \\ &&-2\left[{\cal H}_{20,-18}(x_i)(1+\beta)- {\cal H}_{6}(x_i)(-1 +\beta)\vphantom{\int_0^{x_2}}\right]
\nonumber \\ &&-\frac{1}{M_{B}^{2}}\left[\vphantom{\int_0^{x_2}} \left\{ 2{\cal H}_{20,18}(x_i)(1+\beta)(Q^{2}+s(t_1,Q^2)+m_{N}^{2}(-1+t_{1}))\right\} \right.
\nonumber \\ &&+m_{N}^{2}\left\{{\cal H}_{15,-14}(x_i)t_{1}(1-\beta)-4{\cal H}_{21,24}(x_i)t_{1}(1+\beta)\right.
\nonumber \\ &&\left.+2{\cal H}_{10}(x_i)(-1+\beta)(t_{1}-x_{2})+2({\cal H}_{16}(x_i)(-1 +\beta)+2{\cal H}_{24}(x_i)(1+\beta))x_{2}\right\} 
\vphantom{\int_0^{x_2}}\left] \vphantom{\int_0^{x_2}}\right) 
\nonumber\\
&-&\left.e_{u}m_{N}\int_{t_{0}}^{1}dx_{2}\int_{0}^{1-x_{2}}dx_{1}e^{-s_{0}/M_{B}^{2}}\frac{t_{0}}{Q^{2}+m_{N}^{2}t_{0}^{2}}\right( 2{\cal H}_{20,18}(x_i)(1+\beta)(Q^{2}+s_{0}+m_{N}^{2}(-1+t_{0}))
\nonumber \\ &&+m_{N}^{2}\left[ \vphantom{\int_0^{x_2}}\left\{ {\cal H}_{-8,9}(x_i)(1-\beta)-(3{\cal H}_{21,24}(x_i)+8{\cal H}_{23}(x_i))(1+\beta)\right\} t_{0}\right.
\nonumber \\ && +2{\cal H}_{10}(x_i)(-1+\beta)(t_{0}-x_{2})
+2({\cal H}_{16}(x_i)(-1 +\beta)+{\cal H}_{24}(x_i)(1+\beta))x_{2}\vphantom{\int_0^{x_2}}\left]  \vphantom{\int_0^{x_2}}\right) 
\nonumber \\ &&\left.+e_{d}\eta_{1}'(Q^{2},\beta)+e_{u}\eta_{1}(Q^{2},\beta)\vphantom{\int_0^{x_2}}\right\},\nonumber\\
\end{eqnarray}
\begin{eqnarray}\label{F_{2}}
&&F_{2}(Q^{2})=
\frac{-m_{N}}{\lambda_{N}} \left. e^{m_{N}^{2}/M_{B}^{2}}\right\{e_{u}\int_{t_{0}}^{1}dx_{2}\int_{0}^{1-x_{2}}dx_{1}
e^{-s(x_{2},Q^{2})/M_{B}^{2}}\left[\frac{2{\cal H}_{5}(x_i)(-1+\beta)}{x_{2}}\right](x_i)
\nonumber \\ &&\left.-e_{u}m_{N}^{2}\int_{t_0}^1dx_2\int_0^{1-x_2}dx_1\int_{t_0}^{x_2}\frac{dt_1}{t_1}e^{-s(t_1,Q^2)/M_{B}^{2}
}\right( \nonumber\\
&&\frac{1}{M_{B}^{2}}\left[{\cal H}_{8,-9}(x_i)(1-\beta)+2( {\cal H}_{18,20}(x_i)+2{\cal H}_{21,22}(x_i)+4{\cal H}_{23}(x_i))(1 +\beta)\vphantom{\int_0^{x_2}}\right]
\nonumber \\ &&\left. -\frac{4}{M_{B}^{2}t_{1}}\left[ {\cal H}_{22}(x_i)(1+\beta)x_{2}\vphantom{\int_0^{x_2}}\right]\vphantom{\int_0^{x_2}}\right) 
\nonumber \\ &&e_{u}m_{N}^{2}\int_{t_{0}}^{1}dx_{2}\int_{0}^{1-x_{2}}dx_{1}e^{-s_{0}/M_{B}^{2}}\left(\frac{1}{Q^{2}+m_{N}^{2}t_{0}^{2}}\right[ {\cal H}_{8,-9}(x_i)(-1+\beta)t_{0}
\nonumber \\ &&-2({\cal H}_{18,20}(x_i)+2{\cal H}_{21,22}(x_i)+4{\cal H}_{23}(x_i))(1+\beta)t_{0}+4 {\cal H}_{22}(x_i)(1+\beta)x_{2}
\vphantom{\int_0^{x_2}}\left]\vphantom{\int_0^{x_2}}\right) \nonumber\\
&&\left.+e_{d}\eta_{2}'(Q^{2},\beta)+e_{u}\eta_{2}(Q^{2},\beta)\vphantom{\int_0^{x_2}}\right\},
\end{eqnarray}

where
\begin{eqnarray}
{\cal F}(x_i) &=& {\cal F}(x_1,x_2,1-x_1-x_2),
\nonumber \\
{\cal F}(x_i') &=& {\cal F}(x_1,1-x_1-x_3,x_3),\nonumber \\
s(y,Q^2)&=&(1-y)m_{N}^2+\frac{(1-y)}{y}Q^2,
\end{eqnarray}
with $t_{0}(s_{0},Q^2)$ being solution of the equation $s(t_{0},Q^2)=s_{0}$, 
and
\begin{eqnarray}\label{eta_{1}}
&&\eta_{1}(Q^{2},\beta)=
  m_{N}\left\{
\int_{t_{0}}^{1}dx_{3}\int_{0}^{1-x_{3}}dx_{1}e^{-s(x_{3},Q^{2})/M_{B}^{2}}\right[({\cal H}_{1,17,3}(x_i)-2{\cal H}_{19}(x_i))(1+\beta)\nonumber \\ && \left.+{\cal H}_{13,7}(x_i) (-1+\beta)\vphantom{\int_0^{x_2}}\right]
+\int_{t_0}^1dx_3\int_0^{1-x_3}dx_1\int_{t_0}^{x_3}dt_1e^{-s(t_1,Q^2)/M_{B}^{2}
}\left(\vphantom{\int_0^{x_2}}\right.\nonumber\\
&& \frac{1}{M_{B}^{4}t_{1}}\left[-{\cal H}_{22}(x_i)m_{N}^{2}(-m_{N}^{2}+Q^{2}+s(t_1,Q^2))(1+\beta)x_{3}\vphantom{\int_0^{x_2}}\right]\nonumber\\
&&+\frac{1}{M_{B}^{4}}\left[{\cal H}_{22}(x_i)m_{N}^{2}(m_{N}^{2}(-1+2t_{1}-2x_{3})+Q^{2}+s(t_1,Q^2))(1+\beta)\vphantom{\int_0^{x_2}}\right]\nonumber\\
&&+\frac{1}{2M_{B}^{2}t_{1}}\left[\vphantom{\int_0^{x_2}} -(m_{N}^{2}-Q^{2}-s(t_1,Q^2))\left\{({\cal H}_{18}(x_i)-3{\cal H}_{20}(x_i))(1+\beta)+2{\cal H}_{6,12}(x_i)(-1+\beta)\right\}\right.\nonumber\\
&&\left.+2(2{\cal H}_{22}(x_i)-{\cal H}_{24}(x_i))m_{N}^{2}(1+\beta)x_{3}\vphantom{\int_0^{x_2}}\right]\nonumber\\
&&+\frac{1}{M_{B}^{2}}\left[m_{N}^{2}\left\{{\cal H}_{-12,15,-6,9}(x_i)(1-\beta)+({\cal H}_{18,-2,24,4,21}(x_i)+2{\cal H}_{-20,-22,23}(x_i))(1+\beta)\right\}\vphantom{\int_0^{x_2}}\right]\nonumber\\
&&\left.+\frac{1}{t_{1}}\left[{\cal H}_{12,6}(x_i)(1-\beta)+{\cal H}_{-18,20}(x_i)(1+\beta)\vphantom{\int_0^{x_2}}\right]\vphantom{\int_0^{x_2}}\right)+\int_{t_0}^1dx_3\int_0^{1-x_3}dx_1e^{-s_{0}/M_{B}^{2}}\left(\vphantom{\int_0^{x_2}}\right.\nonumber\\
&&\frac{1}{M_{B}^{2}t_{0}(Q^{2}+m_{N}^{2}t_{0}^{2})}\left[{\cal H}_{22}(x_i)m_{N}^{2}(1+\beta)t_{0}^{2}(Q^{2}+s_{0}+m_{N}^{2}(-1+2t_{0}))(t_{0}-x_{3})\vphantom{\int_0^{x_2}}\right]\nonumber\\
&&+\frac{1}{(Q^{2}+m_{N}^{2}t_{0}^{2})^{3}}\left[2{\cal H}_{22}(x_i)m_{N}^{4}(1+\beta)t_{0}^{4}(Q^{2}+s_{0}+m_{N}^{2}(-1+2t_{0}))(t_{0}-x_{3})\vphantom{\int_0^{x_2}}\right]\nonumber\\
&&-\frac{1}{(Q^{2}+m_{N}^{2}t_{0}^{2})^{2}}\left[{\cal H}_{22}(x_i)m_{N}^{2}(1+\beta)t_{0}^{2}((Q^{2}+s_{0})(2t_{0}-x_{3}) \right.
\nonumber \\ 
&& + \left. m_{N}^{2}(2t_{0}(-1+3t_{0}-2x_{3})+x_{3}))\vphantom{\int_0^{x_2}}\right]\nonumber\\
&&+\frac{1}{t_{0}(Q^{2}+m_{N}^{2}t_{0}^{2})}\left[2{\cal H}_{22}(x_i)m_{N}^{2}(1+\beta)t_{0}^{2}x_{3}\vphantom{\int_0^{x_2}}\right]+\frac{1}{2(Q^{2}+m_{N}^{2}t_{0}^{2})}\left[\vphantom{\int_0^{x_2}} \right. \nonumber\\
&&-{\cal H}_{20}(x_i)(1+\beta)t_{0}\{3(Q^{2}+s_{0})+m_{N}^{2}(-3+4t_{0})\}+2{\cal H}_{6,12}(x_i)(-1+\beta)(Q^{2}+s_{0}\nonumber\\
&&+m_{N}^{2}(-1+t_{0}))t_{0}+2{\cal H}_{24}(x_i)m_{N}^{2}(1+\beta)t_{0}(t_{0}-x_{3})
\nonumber \\ 
&& +2{\cal H}_{18}(x_i)(1+\beta)t_{0}(Q^{2}+s_{0}+m_{N}^{2}(-1+2t_{0}))\nonumber\\
&&\left.+2m_{N}^{2}{\cal H}_{9,-15}(x_i)(-1+\beta)t_{0}^{2}+2m_{N}^{2}({\cal H}_{4,-2,21}(x_i)+2{\cal H}_{23,-22}(x_i))(1+\beta)t_{0}^{2}\vphantom{\int_0^{x_2}}\left]\vphantom{\int_0^{x_2}}\right)\vphantom{\int_0^{x_2}}\right\},\nonumber\\
\end{eqnarray}
\begin{eqnarray}\label{eta_{2}}
&&\eta_{2}(Q^{2},\beta)=
\int_{t_{0}}^{1}dx_{3}\int_{0}^{1-x_{3}}dx_{1}e^{-s(x_{3},Q^{2})/M_{B}^{2}}\left[\frac{{\cal H}_{11,-5}(x_i)(-1+\beta)}{x_{3}}\right]\nonumber\\
&&+
m_{N}\int_{t_0}^1dx_3\int_0^{1-x_3}dx_1\int_{t_0}^{x_3}dt_1e^{-s(t_1,Q^2)/M_{B}^{2}
}\left( \frac{-2}{M_{B}^{4}}\left[{\cal H}_{22}(x_i)m_{N}^{3}(1+\beta)\vphantom{\int_0^{x_2}}\right]\right.\nonumber\\
&&+\frac{1}{M_{B}^{4}t_{1}}\left[{\cal H}_{22}(x_i)m_{N}(1+\beta)(-Q^{2}-s(t_1,Q^2)+m_{N}^{2}(1+2x_{3}))\vphantom{\int_0^{x_2}}\right]\nonumber\\
&&+\frac{1}{M_{B}^{4}t_{1}^{2}}\left[{\cal H}_{22}(x_i)m_{N}(1+\beta)(Q^{2}+s(t_1,Q^2)-m_{N}^{2})x_{3}\vphantom{\int_0^{x_2}}\right]\nonumber\\
&&-\frac{3}{M_{B}^{2}t_{1}^{2}}\left[ {\cal H}_{22}(x_i)m_{N}(1+\beta)x_{3}\vphantom{\int_0^{x_2}}\right]+\frac{m_{N}}{M_{B}^{2}t_{1}}\left[\vphantom{\int_0^{x_2}}{\cal H}_{12,15,6,-9}(x_i)(-1+\beta)\right.\nonumber\\
&&+({\cal H}_{2,-20,-21,-4}(x_i)+3{\cal H}_{22}(x_i)-2{\cal H}_{23}(x_i))(1+\beta)\vphantom{\int_0^{x_2}}\left]\vphantom{\int_0^{x_2}}\right)\nonumber\\
&&+m_{N}\int_{t_0}^1dx_3\int_0^{1-x_3}dx_1e^{-s_{0}/M_{B}^{2}}\left(\vphantom{\int_0^{x_2}}\right.\nonumber\\
&&-\frac{m_{N}}{M_{B}^{2}(Q^{2}+m_{N}^{2}t_{0}^{2})}\left[{\cal H}_{22}(x_i)(1+\beta)(Q^{2}+s_{0}+m_{N}^{2}(-1+2t_{0}))(t_{0}-x_{3})\vphantom{\int_0^{x_2}}\right]\nonumber\\
&&+\frac{1}{(Q^{2}+m_{N}^{2}t_{0}^{2})^{3}}\left[-2{\cal H}_{22}(x_i)m_{N}^{3}(1+\beta)t_{0}^{3}(Q^{2}+s_{0}+m_{N}^{2}(-1+2t_{0}))(t_{0}-x_{3})\vphantom{\int_0^{x_2}}\right]\nonumber\\
&&+\frac{1}{(Q^{2}+m_{N}^{2}t_{0}^{2})^{2}}\left[{\cal H}_{22}(x_i)m_{N}(1+\beta)t_{0}^{2}(Q^{2}+s_{0}+m_{N}^{2}(-1+4t_{0}-2x_{3}))\vphantom{\int_0^{x_2}}\right]\nonumber\\
&&+\frac{m_{N}}{(Q^{2}+m_{N}^{2}t_{0}^{2})}\left[\vphantom{\int_0^{x_2}}{\cal H}_{-12,-15,-6,9}(x_i)(1-\beta)+({\cal H}_{2,-20,-21,22,-4}(x_i)-2{\cal H}_{23}(x_i))(1+\beta)t_{0}\right.\nonumber\\
&&-{\cal H}_{22}(x_i)(1+\beta)(3x_{3}-2t_{0})\vphantom{\int_0^{x_2}}\left]\vphantom{\int_0^{x_2}}\right).\nonumber\\
\end{eqnarray}
and $\eta_i'(Q^2,\beta)$, ($i=1,~2$) are obtained from $\eta_i(Q^2,\beta)$ by replacing $x_3$ with $x_2$ and replacing ${\cal F}(x_i)$ with ${\cal F}(x_i')$ in the integrals. 
In the above equations, we have used  the short hand notations for the functions ${\cal H}_{\pm i,\pm j, ...}=\pm{\cal H}_{i}\pm{\cal H}_{j}...$, and  ${\cal H}_{i}$ are defined in terms of the distribution amplitudes as follows:
\begin{eqnarray}
&&{\cal H}_{1}=S_{1}~~~~~~~~~~~~~~~~~~~~~~~~~~~~~~~~~~~~~~~~{\cal H}_{2}=S_{1,-2}\nonumber\\&&{\cal H}_{3}=P_{1}~~~~~~~~~~~~~~~~~~~~~~~~~~~~~~~~~~~~~~~~{\cal H}_{4}=P_{1,-2}\nonumber\\&&{\cal H}_{5}=V_{1}~~~~~~~~~~~~~~~~~~~~~~~~~~~~~~~~~~~~~~~~{\cal H}_{6}=V_{1,-2,-3}\nonumber\\&&{\cal H}_{7}=V_{3}~~~~~~~~~~~~~~~~~~~~~~~~~~~~~~~~~~~~~~~~{\cal H}_{8}=-2V_{1,-5}+V_{3,4}\nonumber\\&&{\cal H}_{9}=V_{4,-3}~~~~~~~~~~~~~~~~~~~~~~~~~~~~~~~~~~~~~{\cal H}_{10}=-V_{1,-2,-3,-4,-5,6}\nonumber\\&&{\cal H}_{11}=A_{1}~~~~~~~~~~~~~~~~~~~~~~~~~~~~~~~~~~~~~~~{\cal H}_{12}=-A_{1,-2,3}\nonumber\\&&{\cal H}_{13}=A_{3}~~~~~~~~~~~~~~~~~~~~~~~~~~~~~~~~~~~~~~~{\cal H}_{14}=-2A_{1,-5}-A_{3,4}\nonumber\\&&{\cal H}_{15}=A_{3,-4}~~~~~~~~~~~~~~~~~~~~~~~~~~~~~~~~~~~~{\cal H}_{16}=A_{1,-2,3,4,-5,6}\nonumber\\&&{\cal H}_{17}=T_{1}~~~~~~~~~~~~~~~~~~~~~~~~~~~~~~~~~~~~~~~~{\cal H}_{18}=T_{1,2}-2T_{3}\nonumber\\&&{\cal H}_{19}=T_{7}~~~~~~~~~~~~~~~~~~~~~~~~~~~~~~~~~~~~~~~~{\cal H}_{20}=T_{1,-2}-2T_{7}\nonumber\\&&{\cal H}_{21}=-T_{1,-5}+2T_{8}~~~~~~~~~~~~~~~~~~~~~~~~~~{\cal H}_{22}=T_{2,-3,-4,5,7,8}\nonumber \\&&{\cal H}_{23}=T_{7,-8}~~~~~~~~~~~~~~~~~~~~~~~~~~~~~~~~~~~~~{\cal H}_{24}=-T_{1,-2,-5,6}+2T_{7,8},\nonumber \\
\end{eqnarray}
 where for any distribution amplitudes,  $X_{\pm i,\pm j, ...}=\pm X_{i}\pm X_{j}...$ are also used.
The overlap amplitude of the nucleon interpolating current with nucleon
is determined from sum rule and its expression is \cite{Ozpineci}
\begin{eqnarray}\label{resedue}
&&\lambda^2_{N}=e^{m_{N}^{2}/M_{B}^{2}}\left\{\frac{M_{B}^{6}}{256\pi^{4}}E_{2}(x)(5+2\beta+\beta^{2})+
\frac{<\bar{u}u>}{6}\right[ -6(1-\beta^{2})<\bar{d}d> \nonumber\\&& +(-1+\beta)^{2}<\bar{u}u>\left] -
\frac{m_{0}^{2}}{24M_{B}^{2}}<\bar{u}u>\right[ -12(1-\beta^{2})<\bar{d}d> \nonumber\\&&+(-1+\beta)^{2}<\bar{u}u>\left]
\vphantom{\frac{M_{B}^{6}}{256\pi^{4}}}\right\}.
\end{eqnarray}
where $x=s_{0}/M_{B}^{2}$ and the function
\begin{equation}\label{e}
E_{n}(x)=1-e^{-x}\sum_{k=1}^{n}\frac{x^{k}}{k!}
\end{equation}
corresponds to the continuum subtraction.
\section{Numerical results}
It follows from explicit expressions of the sum rules for the nucleon
electromagnetic  form factors that, the nucleon DA's are
the principal input parameters, whose explicit expressions 
can be found in \cite{Lenz}. These DA's contain nonperturbative
parameters which should be determined in some framework. 
In the present work, we consider two different determination
of these input parameters: a) All eight nonperturbative parameters
$f_{N},~\lambda_{1},
~\lambda_{2},~V_{1}^{d},~A_{1}^{u},~f_{1}^{d},~f_{1}^{u}$ and
$f_{2}^{d}$ are estimated within QCD sum rules method \cite{Lenz,
Braun1, Braun2} (set1), b) The condition that the next to leading
conformal spin contributions vanish, fixes five of the eight parameters.
This  is the so called asymptotic set. The values of all nonperturbative
parameters are (see \cite{Lenz}):
\begin{center}
\begin{eqnarray}
f_{N} &=& (5.0\pm0.5)\times10^{-3}~GeV^{2},\nonumber \\
\lambda_{1} &=& -(2.7\pm0.9)\times10^{-2}~GeV^{2}, \nonumber\\
\lambda_{2} &=& (5.4\pm1.9)\times10^{-2}~GeV^{2},\nonumber\\
\end{eqnarray}
\end{center}
\begin{eqnarray}
&&\mbox{set1 }~~~~~~~~~~~~~~~~~~~~~~~~~~~~~~~~~~~~~~~~~~~~~~~~~~~~~~\mbox{asymptotic }\nonumber\\
 A_{1}^{u} &=& 0.38\pm0.15,~~~~~~~~~~~~~~~~~~~~~~~~~~~~~ ~~~~~~~~~~~~~~~~~A_{1}^{u} = 0,\nonumber \\
V_{1}^{d} &=& 0.23\pm0.03,~~~~~~~~~~~~~~~~~~~~~~~~~~~~~~~~~~~~~~~~~~~~~~V_{1}^{d} = \frac{1}{3}, \nonumber\\
 f_{1}^{d} &=&0.40\pm0.05,~~~~~~~~~~~~~~~~~~~~~~~~~~~~~~~~~~~~~~~~~~~~~~f_{1}^{d} = \frac{3}{10}, \nonumber \\
  f_{2}^{d} &=&0.22\pm0.05,~~~~~~~~~~~~~~~~~~~~~~~~~~~~~~~~~~~~~~~~~~~~~~f_{2}^{d} = \frac{4}{15},\nonumber \\
   f_{1}^{u} &=& 0.07\pm0.05,~~~~~~~~~~~~~~~~~~~~~~~~~~~~~~~~~~~~~~~~~~~~~~f_{1}^{u} = \frac{1}{10}.
\end{eqnarray}

The continuum threshold that appears in the continuum subtraction is determined from the 
mass sum rules as $s_0=2.25~GeV^2$. There are two auxiliary parameters of the sum
rules: the Borel parameter
$M_{B}^{2}$ and the parameter $\beta$. 
The Borel mass square $M_{B}^{2}$ is the artificial parameter of the
sum rules and therefore wee need to find a region of $M_{B}^{2}$,
where physically measurable quantities, in our case electromagnetic
form factors, be independent of $M_{B}^{2}$. Lower bound of
$M_{B}^{2}$ is determined from condition that contribution from higher states and
continuum in the correlator should be enough small, upper bound of
$M_{B}^{2}$ is determined from condition that series of the light
cone expansion with increasing twist should be convergent. Our
numerical analysis shows that both conditions are satisfied in the
region $1 GeV^{2}\leq M_{B}^{2}\leq 2GeV^{2}$, which we will use in
numerical analysis.

The other auxiliary parameter $\beta$ is chosen in a region
such that, the predictions are independent of the precise value of $\beta$ in that region.
In our analysis, it is shown that in the region $-0.5\leq
cos\theta\leq 0.5$ the form factors are practically 
insensitive to the variation of $\beta$, where $\theta$ is defined as
$tan\theta=\beta$. Note that the analysis of mass
sum rules and magnetic moments of octet baryons \cite{Ozpineci}
leads to the very close region for $cos\theta$, i.e. $-0.6\leq
cos\theta\leq 0.3$. Also, it is observed in \cite{Leinweber}  that
the optimal value of $\beta$ is $\beta=-1.2$($\cos \theta = - 0.64$), which follows from the Monte Carlo
analysis.

\begin{figure}[h!]
\begin{center}
\includegraphics[width=13cm]{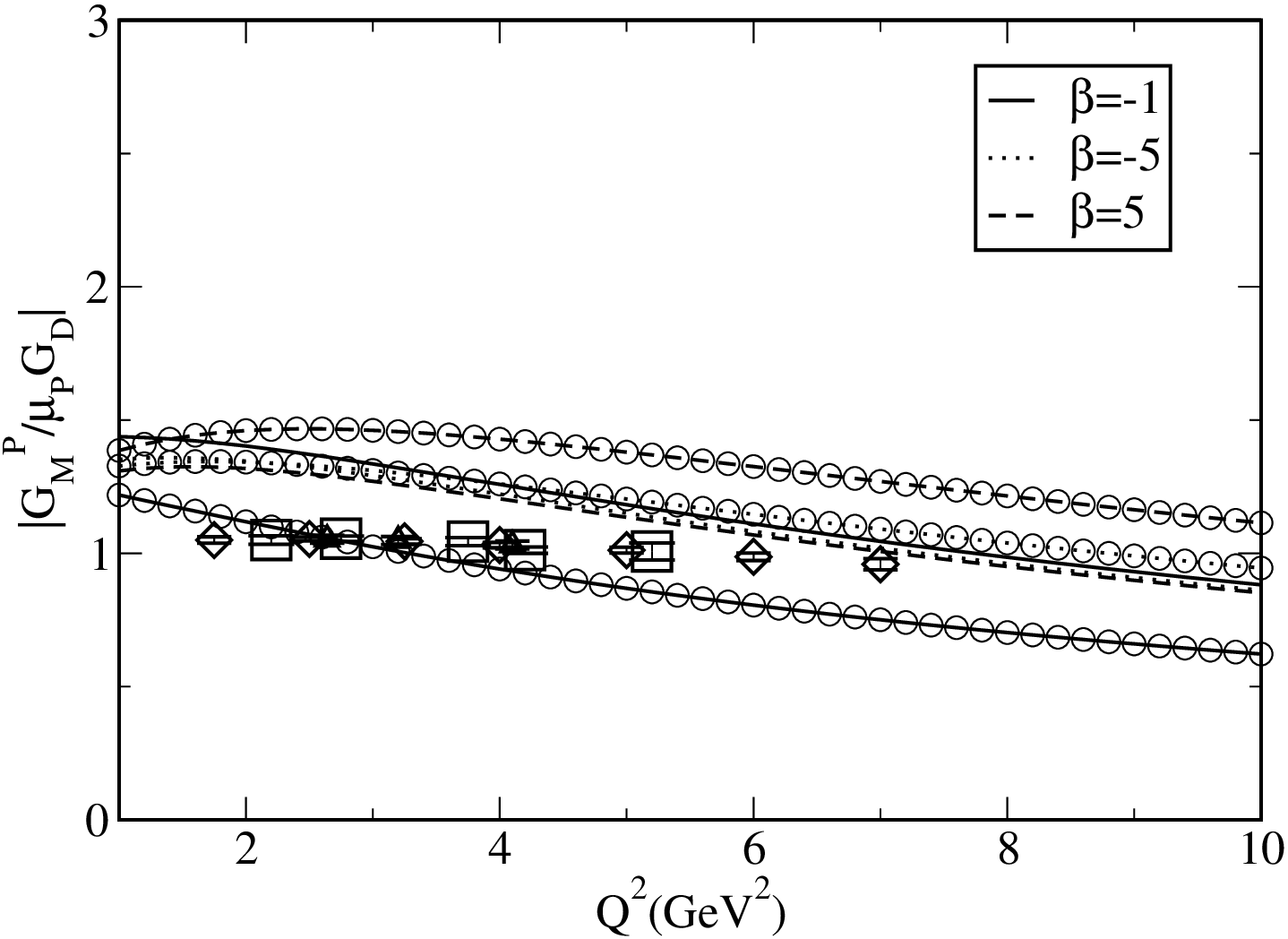}
\end{center}
\caption{The dependence of $G^{P}_{M}/\mu_{P}G_{D}$ on $Q^{2}$ at
$s_{0}=2.25~GeV^{2}, M_{B}^{2}=1.2~GeV^{2}$ for $\beta=-1,~-5$ and $5$. The boxes correspond to 
experimental data in \cite{Christy}, the diamonds  to \cite{Andivahis} and the up-triangles to
\cite{Qattan} . The lines with circles correspond to set1 and the lines without any circles correspond
to the asymptotic DA's} 
\label{fig1}
\end{figure}

\begin{figure}[h!]
\begin{center}
\includegraphics[width=13cm]{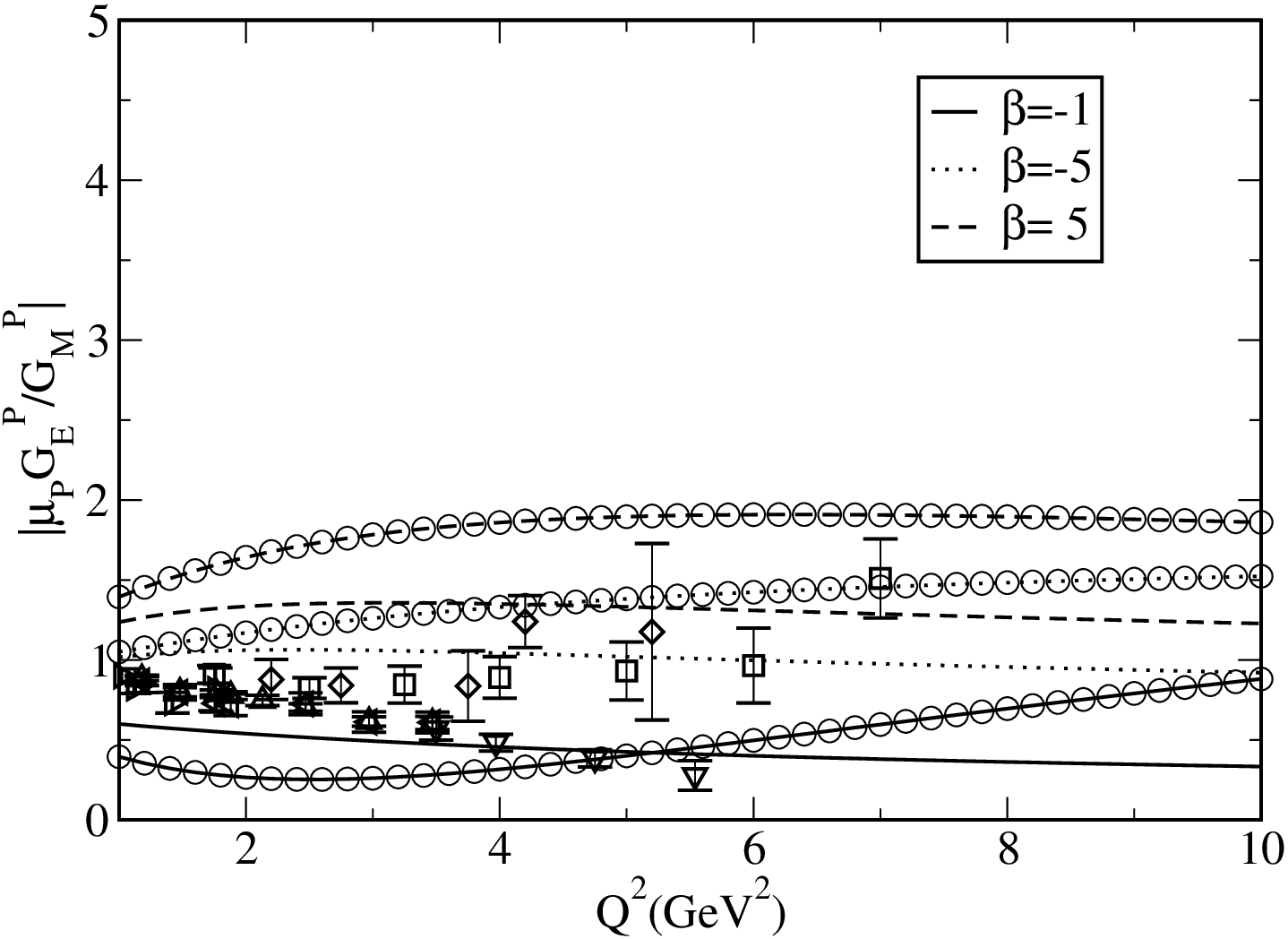}
\end{center}
\caption{
The same as Fig. \ref{fig1}, but for $\mu_{P}G^{P}_{E}/G^{P}_{M}$.
The boxes/diamonds/up-triangles/down-triangles/right-triangles/left-triangles correspond to experimental data given in \cite{Christy}/\cite{Andivahis}/\cite{Punjabi}/\cite{Gayou}/\cite{Gayou2}/\cite{Jones} respectively}
\label{fig2}
\end{figure}

In Fig. \ref{fig1}, we present the dependence of the proton magnetic form
factor $G^{p}_{M}/\mu_{p}G_{D}$ on $Q^{2}$ at $s_{0}=2.25~GeV^{2}$,
$M_{B}^{2}=1.2~GeV^{2}$ for two sets of DA's, at fixed values of
parameter $\beta$. In this figure, we also present the experimental
results \cite{Christy, Andivahis, Qattan}. From this figure, we see that the $Q^{2}$ dependencies, as
well as the magnitude of proton magnetic form factor are rather in
good agreement with the experimental data, especially for the set 1 of
DA's and Ioffe current ($\beta=-1$). The dependence of the ratio of the
proton electric form factor to the magnetic form factor
$\mu_{p}G^{p}_{E}/G^{p}_{M}$ on $Q^{2}$ at $s_{0}=2.25~GeV^{2}$,
$M_{B}^{2}=1.2~GeV^{2}$ for two sets of DA's, at fixed values of
parameter $\beta$ is depicted in Fig. \ref{fig2}. From this figure it follows
that, practically, both sets of DA's well describe the existing
experimental results, except for $\beta=5$ and  $\beta=-1$ of set 1. For large values of $Q^2$, $Q^2 > 4~GeV^2$, the 
experimental results obtained in \cite{Gayou} and in \cite{Andivahis, Qattan} are not in agreement. Whereas $\beta=-1$ 
describes better the data in \cite{Gayou}, larger values of $\vert \beta \vert$ describe better the data in \cite{Qattan}.

\begin{figure}[h!]
\begin{center}
\includegraphics[width=13cm]{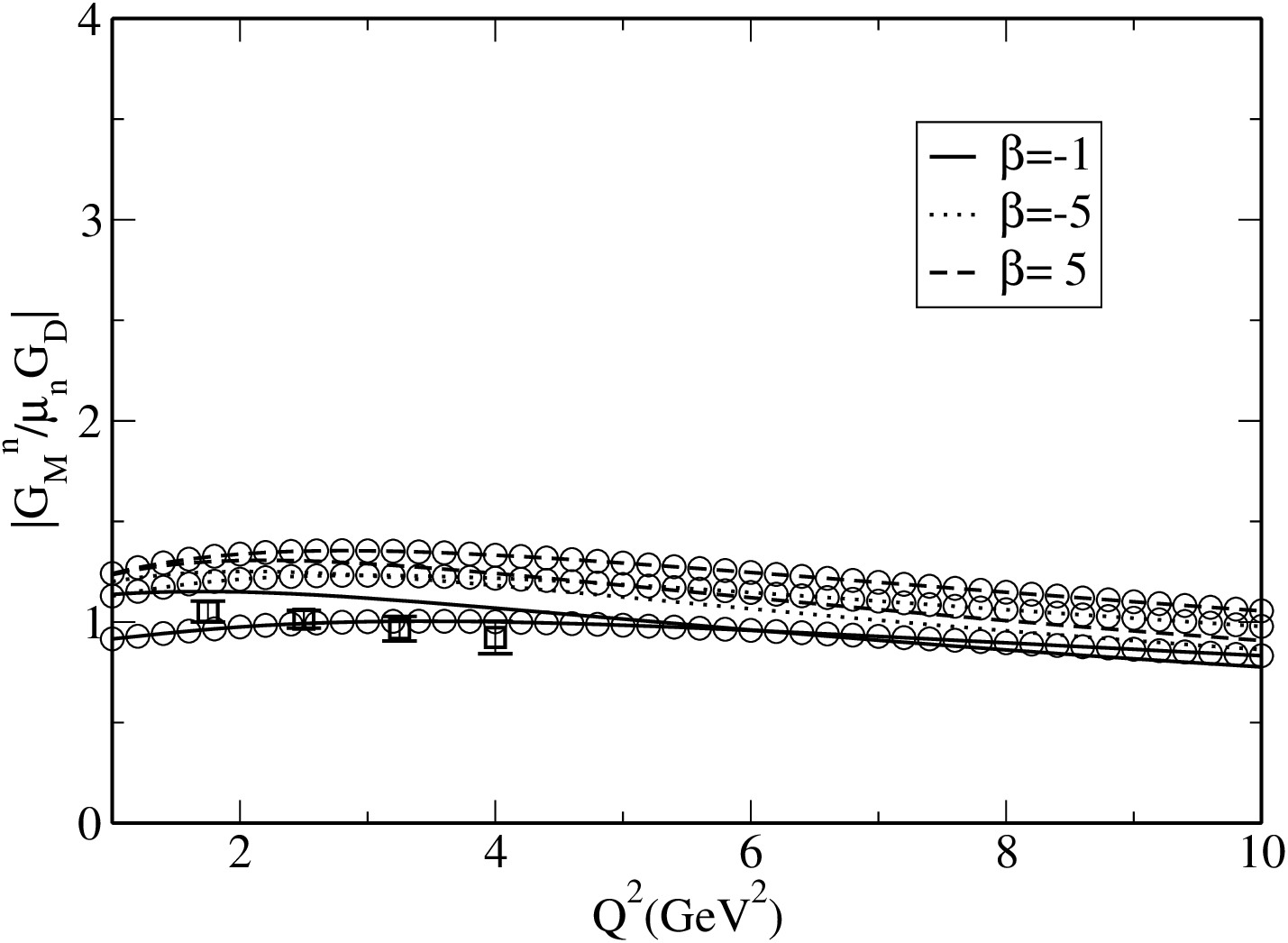}
\end{center}
\caption{The same as Fig. \ref{fig1} but for $G^{n}_{M}/\mu_{n}G_{D}$.
The boxes 
correspond to experimental data (\cite{Lung})} 
\label{fig3} 
\end{figure}

\begin{figure}[h!]
\begin{center}
\includegraphics[width=13cm]{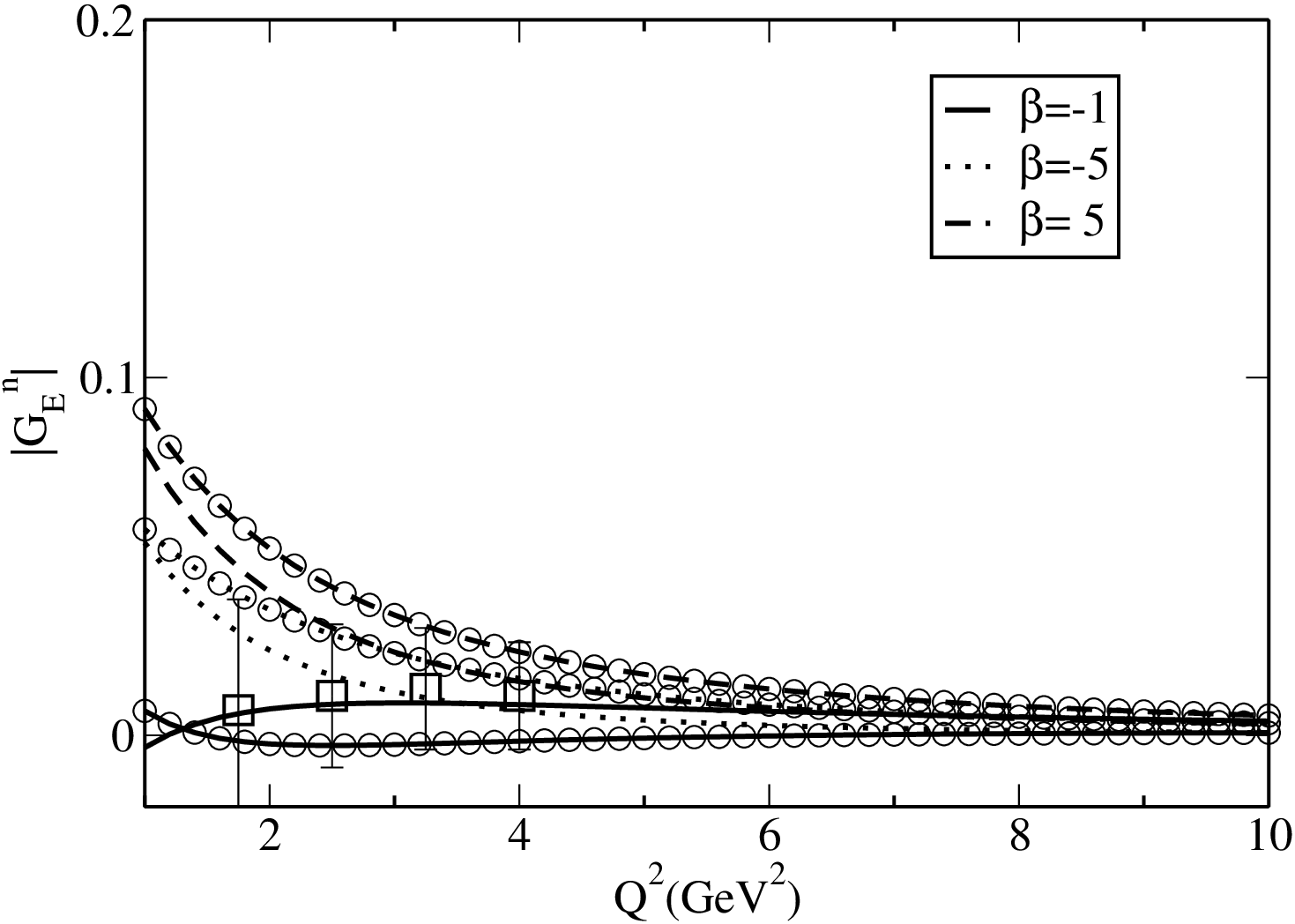}
\end{center}
\caption{The same as Fig. \ref{fig1} but for $G^{n}_{E}$.
 The  boxes are correspond to experimental data
(\cite{Lung})} 
\label{fig4}
\end{figure}

The LCQSR results for the neutron magnetic (normalized to the dipole
form factor ) and electric form factors  are given in Fig. \ref{fig3} and
Fig. \ref{fig4}, respectively. From Fig. \ref{fig3}, we see that the magnetic form factor
of neutron reproduce experimental data very well at $\beta=-1$ for both
sets of DA's. Neutron electric form factor is in a good agreement
with the experimental result for all cases.

Analysis of the experimental results (for review see
\cite{Perdrisat} and references therein) lead that the magnetic form
factors of the nucleon are very well described by the dipole formula
\begin{equation}\label{dipole}
G^{n,p}_{M}(Q^{2})=\frac{\mu_{n,p}}{\left(1+\frac{Q^{2}}{(0.71~GeV)^{2}}\right)^{2}}=\mu_{n,p}G_{D}.
\end{equation}
The measured values of the electric form factors of the neutron are
given in \cite{Zhon,Rohe}. 

\begin{figure}[h!]
\begin{center}
\includegraphics[width=13cm]{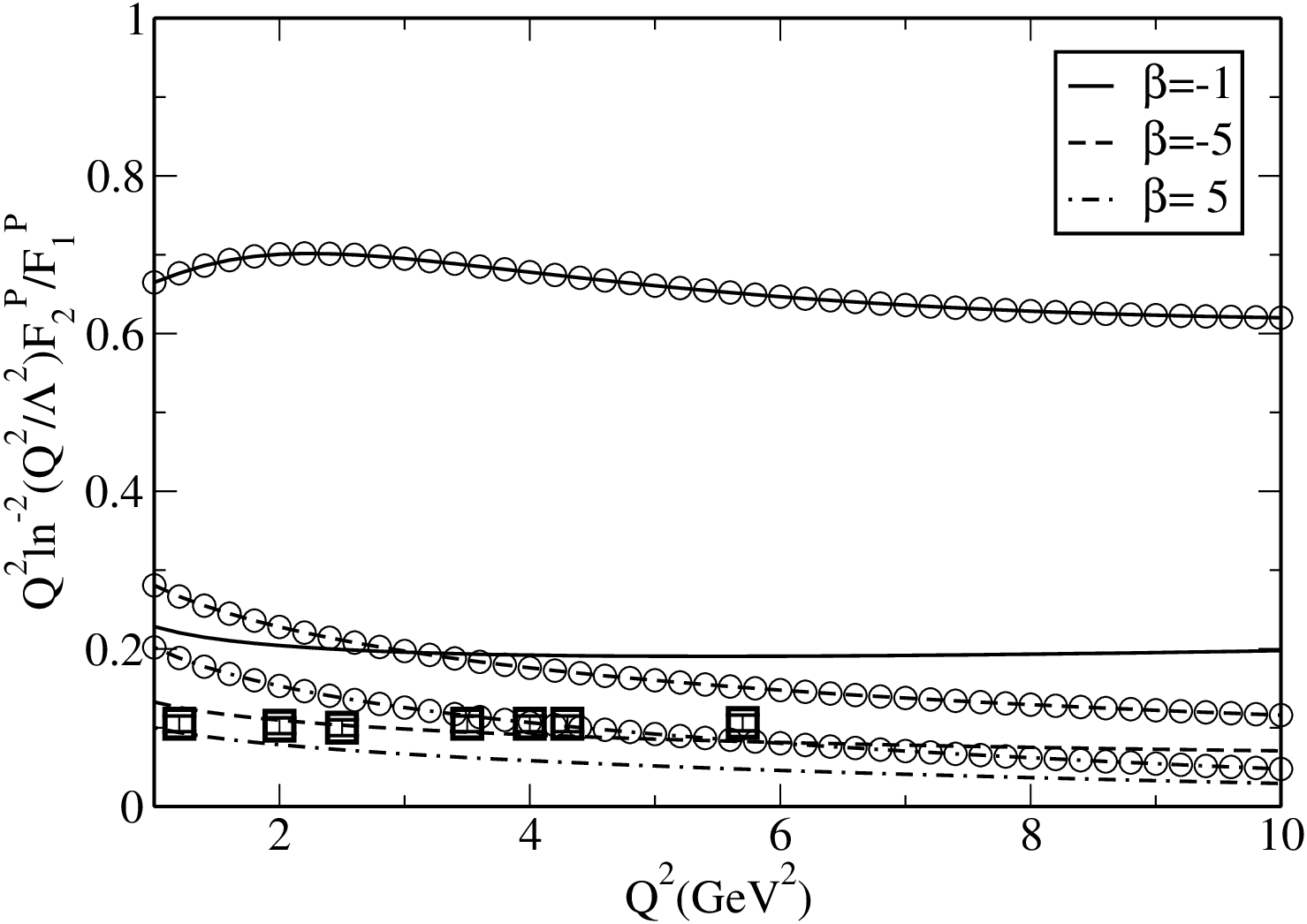}
\end{center}
\caption{The same as Fig. \ref{fig1} but for $Q^{2}ln^{-2}(\frac{Q^{2}}{\Lambda^{2}})F_{2}^{p}/F_{1}^{p}$
 where $\Lambda=300MeV$. The boxes 
correspond to experimental data (\cite{Jager})} 
\label{fig6}
\end{figure}

\begin{figure}[h!]
\begin{center}
\includegraphics[width=13cm]{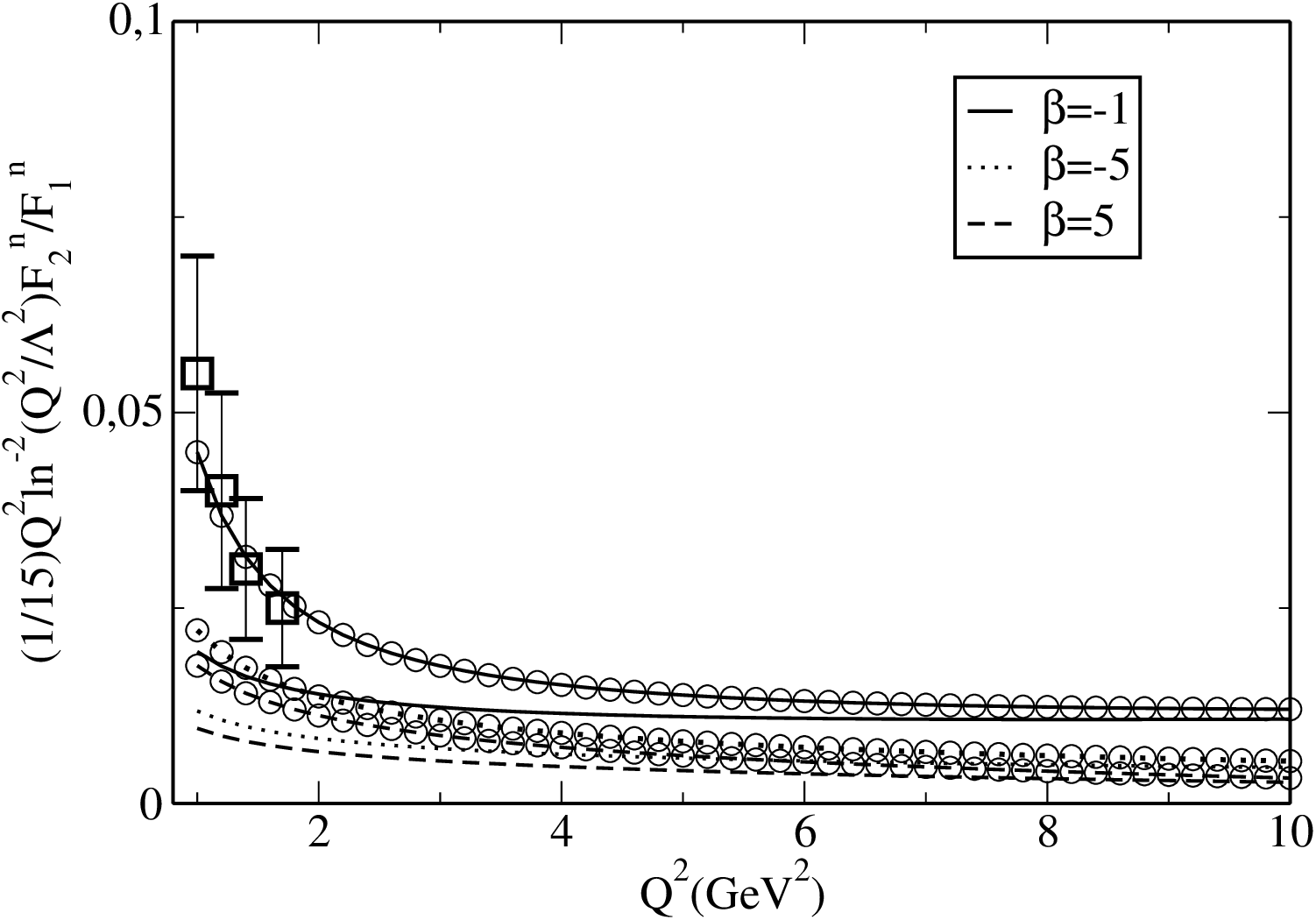}
\end{center}
\caption{The dependence of $\frac{1}{15}Q^{2}ln^{-2}(\frac{Q^{2}}{\Lambda^{2}})F_{2}^{n}/F_{1}^{n}$
on $Q^{2}$ at $s_{0}=2.25~GeV^{2},
M_{B}^{2}=1.2~GeV^{2},\Lambda=300MeV$. The boxes correspond to
experimental data (\cite{Jager})} 
\label{fig7}
\end{figure}

In \cite{Brodsky,Belitsky}, the following large $Q^{2}$ behavior of
the electromagnetic form factors is obtained
\begin{equation}\label{beh}
\frac{F_{2}(Q^{2})}{F_{1}(Q^{2})}\sim\frac{ln^{2}(Q^{2}/\Lambda^{2})}{Q^{2}}
\end{equation}
where $\Lambda=300~MeV$. In Fig. \ref{fig6} (\ref{fig7}), we
present the logarithmic scale prediction, i.e.
 ($1/15)ln^{-2}(Q^{2}/\Lambda^{2})Q^{2}F_{2}(Q^{2})/F_{1}(Q^{2})$
for the proton (neutron), with available experimental data \cite{Jager} at fixed
values of $\beta$ for two sets of DA's. From these figures, we see that our
prediction for the proton for
$ln^{-2}(Q^{2}/\Lambda^{2})Q^{2}F_{2}(Q^{2})/F_{1}(Q^{2})$
is in good agreement with experimental data except for $\beta=-1$ case
for both DA's, and $\beta=-5$ case for set1. For the neutron case only
set1 for $\beta=-1$ describes quite successfully the existing experimental
data.

\begin{figure}[h!]
\begin{center}
\includegraphics[width=13cm]{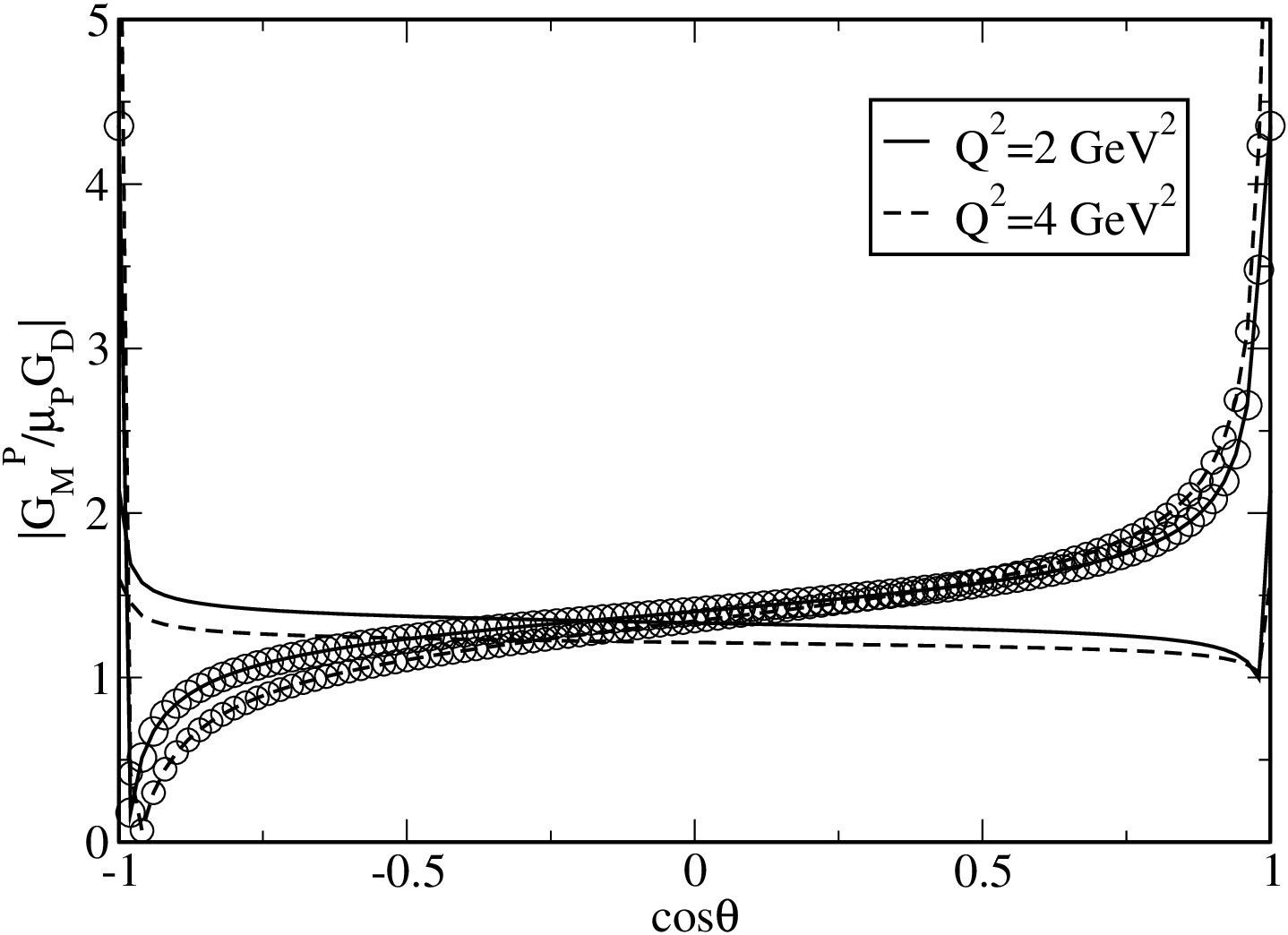}
\end{center}
\caption{The dependence of $G^{P}_{M}/\mu_{P}G_{D}$ on $cos\theta$ at
$s_{0}=2.25~GeV^{2}, M_{B}^{2}=1.2~GeV^{2}$ for two different values of $Q^2$, i.e. $Q^2=2~GeV^2$ and $Q^2=4~GeV^2$. The
lines with circles correspond to set1 and the lines without any circles correspond to the asymptotic wavefunctions} 
\label{fig5}
\end{figure}

Finally, in Fig. \ref{fig5}, as an example on the dependence of the
predictions on $\beta$, 
we present the dependence of proton magnetic form
factor normalized to the dipole form factor $G^{p}_{M}/\mu_{p}G_{D}$
on $cos\theta$, for both sets of DA's at two fixed values of $Q^{2}$.
It follows from this graph that, in the chosen region of $\beta$, i.e. in the region $-0.5\leq
cos\theta\leq 0.5$, the form factor $G^{p}_{M}$ is practically 
insensitive to the variation of $\beta$. 


In conclusion, in present work, we calculate the nucleon
electromagnetic form factors using the most general form of the
nucleon interpolating current in the light cone QCD sum rules. The
sum rules for these form factors are obtained. Using two forms of
the DA's, we calculate sum rules predictions for these form factors
and compare them with existing experimental data. We obtain that
our results are in a good agreement with the existing experimental data.
More precisely, at different values of $\beta$, our results for the form
factors reproduce the experimental data. Finally, we obtained the
``working region for $\beta$''.

Our final remark is that in order to answer to the question which $\beta$ is more preferable,
both theoretical and experimental studies have to be refined. From theoretical part ${\cal O}(\alpha_s)$ corrections
to the distributions amplitudes and more accurate determination of the DA's are needed. From experimental data, the discrepencies
between various data has to be eliminated.

\section{Acknowledgment}
Two of the authors  (K. A. and A. O.), would like to thank TUBITAK, Turkish Scientific and Research Council, for their partial financial support both through the scholarship program 
and also through the project number 106T333. We thank A. Lenz for stimulating discussion and providing us with experimental data.
 
\newpage
\newpage
\newpage
\newpage
\newpage
\newpage
\newpage
\clearpage
\newpage
\end{document}